\documentclass[12pt]{article}

\usepackage{graphicx, amsmath, amssymb, afterpage}

\usepackage{times}

\usepackage{lineno}
\newcommand*\patchAmsMathEnvironmentForLineno[1]{%
  \expandafter\let\csname old#1\expandafter\endcsname\csname #1\endcsname
  \expandafter\let\csname oldend#1\expandafter\endcsname\csname end#1\endcsname
  \renewenvironment{#1}%
     {\linenomath\csname old#1\endcsname}%
     {\csname oldend#1\endcsname\endlinenomath}}%
\newcommand*\patchBothAmsMathEnvironmentsForLineno[1]{%
  \patchAmsMathEnvironmentForLineno{#1}%
  \patchAmsMathEnvironmentForLineno{#1*}}%
\AtBeginDocument{%
\patchBothAmsMathEnvironmentsForLineno{equation}%
\patchBothAmsMathEnvironmentsForLineno{align}%
\patchBothAmsMathEnvironmentsForLineno{flalign}%
\patchBothAmsMathEnvironmentsForLineno{alignat}%
\patchBothAmsMathEnvironmentsForLineno{gather}%
\patchBothAmsMathEnvironmentsForLineno{multline}%
}

\newenvironment{sciabstract}{%
\begin{quote} \bf}
{\end{quote}}

\newcounter{lastnote}

\title{Fault-tolerant quantum error detection}

\author{N. M. Linke,$^{1\ast}$ M. Gutierrez,$^{2}$ K. A. Landsman,$^{1}$ C. Figgatt,$^{1}$
\\ S. Debnath,$^{1}$ K. R. Brown,$^{2}$ C. Monroe$^{1,3}$\\
\\
\normalsize{$^{1}$Joint Quantum Institute, Department of Physics, and}\\
\normalsize{Joint Center for Quantum Information and Computer Science,}\\
\normalsize{University of Maryland, College Park, MD 20742, USA}\\
\normalsize{$^{2}$Schools of Chemistry and Biochemistry,}\\
\normalsize{Computational Science and Engineering, and Physics,}\\
\normalsize{Georgia Institute of Technology, Atlanta, GA 30332, USA}\\
\normalsize{$^{3}$IonQ Inc., College Park, MD 20742, USA}\\
\normalsize{$^\ast$To whom correspondence should be addressed; E-mail:  linke@umd.edu.}
}

\date{}

\begin{document} 


\baselineskip24pt


\maketitle 

\textit{One sentence summary:} We show the fault-tolerant encoding, measurement, and operation of a logical qubit realized in four physical trapped ion qubits, and demonstrate its robustness against intrinsic system errors as well as artificially added errors when compared to a non-fault tolerant logical gauge qubit and a bare physical qubit.


\begin{sciabstract}
Quantum computers will eventually reach a size at which quantum error correction becomes imperative. Quantum information can be protected from qubit imperfections and flawed control operations by encoding a single logical qubit in multiple physical qubits. This redundancy allows the extraction of error syndromes and the subsequent detection or correction of errors without destroying the logical state itself through direct measurement. Here we show the encoding and syndrome measurement of a fault-tolerant logical qubit via an error detection protocol on four physical qubits, represented by trapped atomic ions. This demonstrates for the first time the robustness of a fault-tolerant qubit to imperfections in the very operations used to encode it. The advantage persists in the face of large added error rates and experimental calibration errors.
\end{sciabstract}

\begin{figure}[ht]
\centering
\includegraphics[width=0.5\columnwidth]{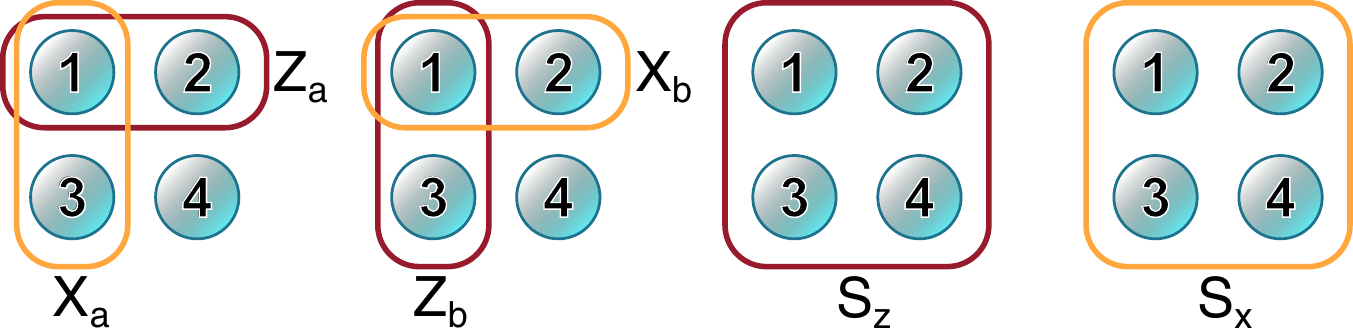} 
\caption{Graphical representation of the logical operators and stabilizers defining the [[4,2,2]] code on physical qubits $1$-$4$. The structure of the logical operators $X$ and $Z$ for the two encoded qubits $L_a$ and $L_b$, and for the two stabilizers $S_x$ and $S_z$, is defined in equations \ref{eq:operators} and \ref{eq:stabilizers}.}
\label{fig:figure1}
\end{figure}

The discovery of quantum error correction (QEC) codes gave credibility to the idea of scaling up physical quantum systems to arbitrary sizes \cite{Calderbank96, Steane96, Preskill98}. Showing that all elements of error correction can be realized in a fault-tolerant way is therefore of fundamental interest. Fault tolerance removes the assumption of perfect encoding and decoding of logical qubits \cite{Nielsen11}, since the logical error probability scales as a convex function of the physical error probability for small errors \cite{Gottesman09}. While several experiments have shown a reduction of high intrinsic or artificially introduced errors in logical qubits \cite{Chiaverini04, Schindler11, Lanyon13, Waldherr14, Nigg14, Kelly15, Corcoles2015, Cramer16, Ofek16}, fault-tolerant encoding of a logical qubit has never been demonstrated. 

Here we implement a four qubit error detection code with two stabilizers (see figure \ref{fig:figure1}). This leaves two possible encoded qubits, $L_a$ and $L_b$, for which errors can be detected: a $[[4,2,2]]$ code \cite{Grassl97, Gottesman16}. The preparation and error detection procedures considered here are fault-tolerant on only a single encoded qubit. From a fault-tolerance perspective, this is a $[[4,1,2]]$ subsystem code where the logical qubit $L_a$ is protected and the gauge qubit $L_b$ is not. As such, the code was used in experiments with photonic qubits \cite{Lu08, Bell14}. By instead considering errors on both encoded qubits, we highlight the importance of fault-tolerance for reducing intrinsic errors and hook errors. The non-fault-tolerant procedures that generate $L_b$ still succeed in reducing added errors. 

The code implements $L_a$ and $L_b$ on only four physical qubits and hence violates the quantum Hamming bound \cite{Gottesman09}, which means that detected errors cannot be uniquely identified and corrected. We must therefore rely on post-selection to find and discard cases where an error occurred. The code does have the advantage of requiring only five physical qubits for the fault-tolerant encoding of $L_a$: four data qubits and one ancilla qubit. 

\begin{figure}
\centering
\includegraphics[width=0.5\columnwidth]{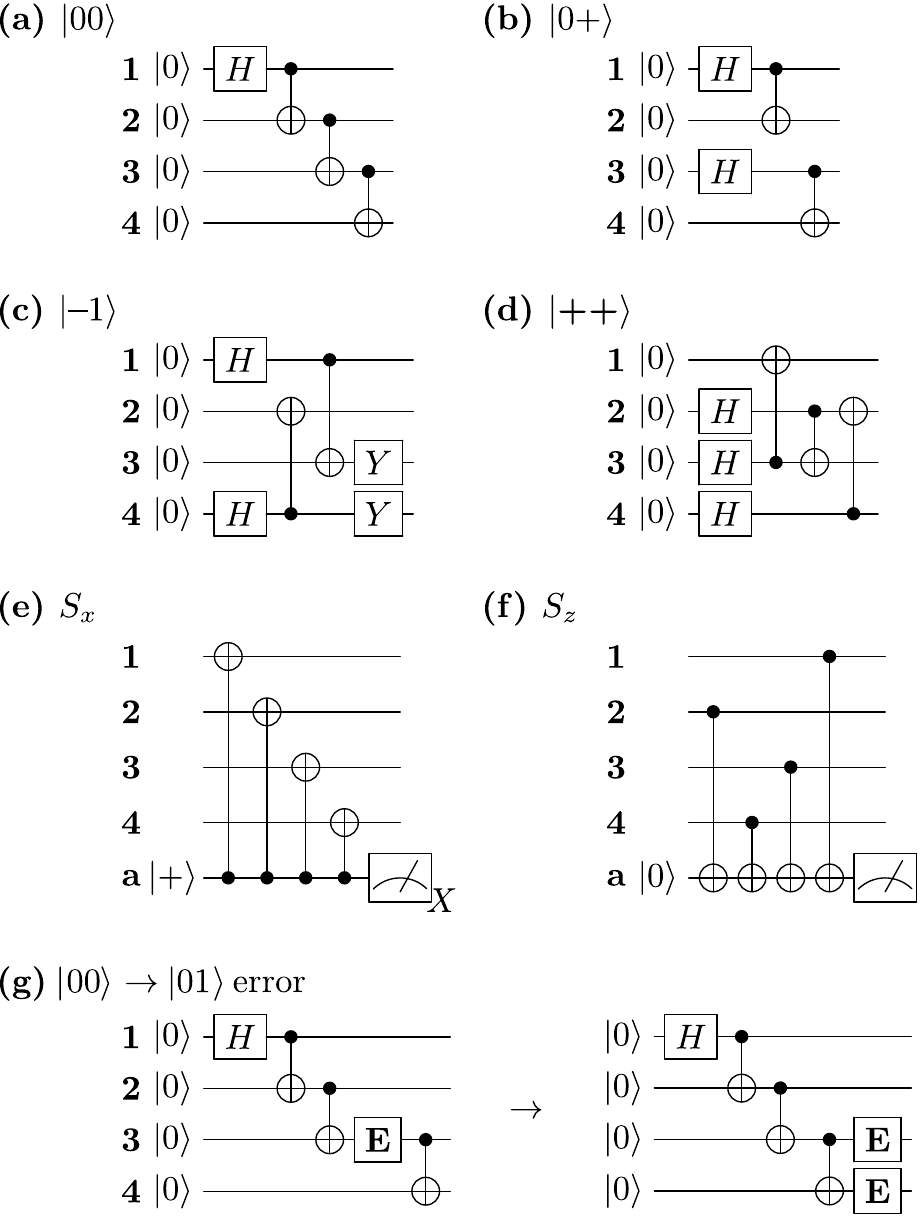} 
\caption{{\bf a-d}: Circuits for the encoding of four different logical states constructed such that logical qubit $L_a$ is fault-tolerant. Any logical state can be achieved by applying single logical qubit operators to states encoded as shown here. {\bf e, f}: Circuits for the two stabilizers $S_x$ and $S_z$, which project $Z-$ and $X-$type errors, respectively, onto an ancilla qubit $a$. Note that a controlled-$Z$ gate is realized by an inverted CNOT with the ancilla in the $Z$-basis as the target. {\bf g}: Example of fault-tolerant construction of circuits for logical qubit $L_a$: The encoding circuit for $|00\rangle_L$ has a single non-detectable error channel. A bit-flip error {\textbf E} occuring as shown can change the state to $|01\rangle_L$, which is an error on the logical gauge qubit $L_b$. Logical qubit $L_a$ is fault-tolerant. This property holds for all circuits {\bf a-f}.}
\label{fig:circuits}
\end{figure}

The logical codewords $|L_a L_b\rangle_L$ in the computational or $z$-basis are
\begin{subequations}
\label{eq:logicalstates}
\begin{align}
|00\rangle_{L} &= (|0000\rangle + |1111\rangle)/\sqrt{2} \label{eq:00} \\
|01\rangle_{L} &= (|0011\rangle + |1100\rangle)/\sqrt{2} \label{eq:01} \\
|10\rangle_{L} &= (|0101\rangle + |1010\rangle)/\sqrt{2} \label{eq:10} \\
|11\rangle_{L} &= (|0110\rangle + |1001\rangle)/\sqrt{2} \label{eq:11},
\end{align}
\end{subequations}
and with $|\pm\rangle=(|0\rangle\pm|1\rangle)/\sqrt{2}$ we can write them down along $x$ as follows:
\begin{subequations}
\label{eq:logicalstatesx}
\begin{align}
|++\rangle_{L} &= (|++++\rangle + |----\rangle)/\sqrt{2} \label{eq:pp}\\
|+-\rangle_{L} &= (|+-+-\rangle + |-+-+\rangle)/\sqrt{2} \label{eq:pm}\\
|-+\rangle_{L} &= (|++--\rangle + |--++\rangle)/\sqrt{2} \label{eq:mp}\\
|--\rangle_{L} &= (|+--+\rangle + |-++-\rangle)/\sqrt{2} \label{eq:mm}.
\end{align}
\end{subequations}
The encoding of different initial states is shown in figure \ref{fig:circuits} (a-d). The fault tolerance arises because the circuits for encoding and syndrome extraction are carefully constructed such that a single physical qubit error occurring anywhere cannot lead to an undetectable error on logical qubit $L_a$. It comes at the cost of the logical gauge qubit $L_b$, for which there is such an undetectable error channel. An example of this is shown in figure \ref{fig:circuits}(g). 

With the Pauli operators $X$, $Y$, $Z$, and the identity $I$, the logical operators are
\begin{subequations}
\label{eq:operators}
\begin{align}
Z_a &=Z \otimes Z \otimes I \otimes I \label{eq:Za}\\
Z_b &=Z \otimes I \otimes Z \otimes I \label{eq:Zb}\\
X_a &=X \otimes I \otimes X \otimes I \label{eq:Xa}\\
X_b &=X \otimes X \otimes I \otimes I \label{eq:Xb}.
\end{align}
\end{subequations}
With these operators and the circuits given in figure \ref{fig:circuits} (a-d), any state $|L_a L_b\rangle_{L}$ can be generated maintaning the fault-tolerance of $L_a$.

The $[[4,2,2]]$ code has the additional advantage that, in contrast to other codes \cite{Cross09}, fault-tolerant syndrome extraction for the logical qubit $L_a$ can be achieved using a bare ancilla, i.e. an ancilla qubit that is not itself a logical qubit. The stabilizers to extract logical phase-flip (Z) and bit-flip (X) errors are $S_x$ and $S_z$, respectively:
\begin{subequations}
\label{eq:stabilizers}
\begin{align}
S_x &=X \otimes X \otimes X \otimes X \label{eq:Sx}\\
S_z &=Z \otimes Z \otimes Z \otimes Z \label{eq:Sz}.
\end{align}
\end{subequations}

As in a Bacon-Shor code block \cite{Bacon06,Shor95}, the code space together with the logical operators and stabilizers form a subsystem that allows local syndrome extraction similar to \cite{Napp13} as depicted in figure \ref{fig:figure1}. The difference is that the stabilizers have weight $4$ since we simultaneously extract information about the gauge qubit $L_b$. Applying these stabilizers conditional on the state of an ancilla qubit extracts the parity of the data qubits along $x$ or $z$ (see figure \ref{fig:circuits}(e,f)). Measuring the ancilla yields either $|0\rangle$, indicating no error, or $|1\rangle$, meaning an error has occurred and the run is to be discarded. With only one ancilla qubit available, we measure the two stabilizers in separate experiments. Since we prepare logical Pauli states, only logical Pauli operations that change the ideal state result in errors. Both stabilizer measurements serve to determine the overall yield, i.e. the fraction of runs for which no error was indicated. In addition to the error checks provided by stabilizer measurements, only even parity outcomes are accepted when the data qubits are measured at the end of the circuit. We note that similar weight-$4$ stabilizers have recently been implemented in superconducting qubits \cite{Takita16}.

We implement the $[[4,2,2]]$ code on a fully-connected quantum computer comprising a chain of five single $^{171}$Yb$^+$ ions confined in a Paul trap (see Methods). The state-detection fidelity for a single qubit is $99.7(1)\%$ for state $|0\rangle$, and $99.1(1)\%$ for state $|1\rangle$. A general $5$-qubit state is detected with $95.7(1)\%$ fidelity. Single- and two-qubit gate fidelities are typically $99.1(5)\%$ and $97(1)\%$, respectively. Typical gate times are $20\:\mu$s for single- and $250\:\mu$s for two-qubit gates. The computational gates H and CNOT are generated by combining several physical-level single- and two-qubit gates in a modular fashion \cite{Debnath16}. 

\begin{figure*}[t]
\centering
\includegraphics[width=0.99\textwidth]{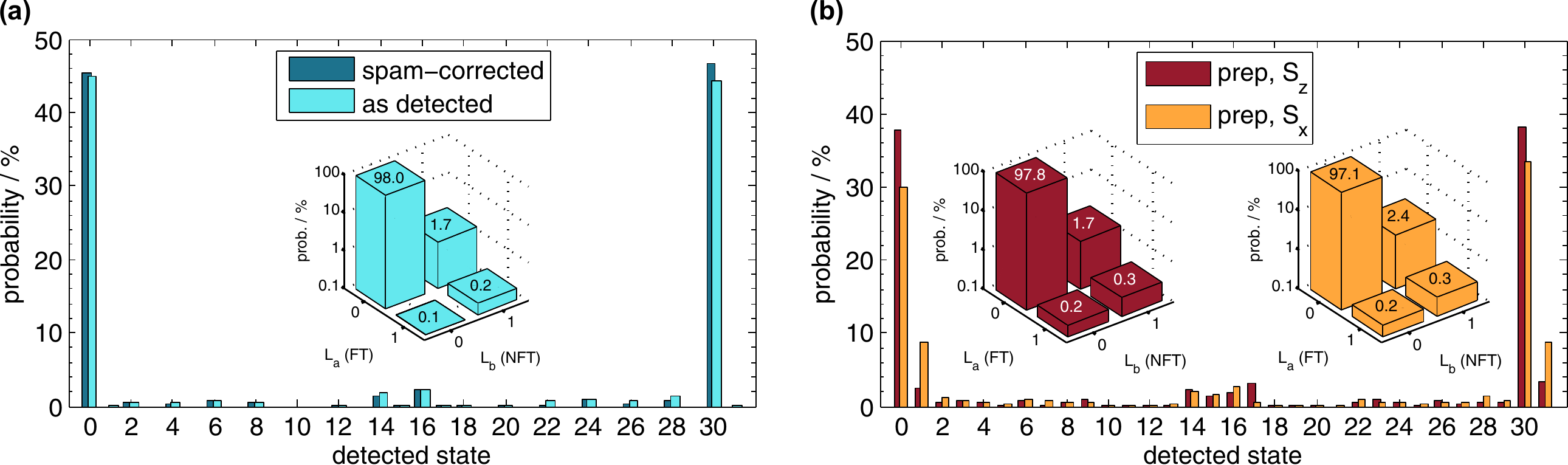}
\caption{(a) Results from the preparation of state $|00\rangle_L$ shown both as detected and processed to correct for physical state preparation and measurement (SPAM) errors. The abscissa represents the $5$-qubit states in decimal. The SPAM-corrected data shows that we succeed in preparing the state with $\sim 92\%$ probability. The inset shows the uncorrected result after post-selection on the state being in the logical basis, i.e. even parity. It is broken down by logical state of the fault-tolerant (FT) qubit $L_a$ and the non-fault-tolerant (NFT) qubit $L_b$. 
(b) Results of the stabilizer measurements after preparation of $|00\rangle_L$. The yields are $77.8\%$ and $65.2\%$ for $S_z$ and $S_x$, respectively. The insets show that the error probability on the fault-tolerant (FT) logical qubit $L_a$ is an order of magnitude below the non-fault-tolerant (NFT) qubit $L_b$.}
\label{fig:prepzero}
\end{figure*}
We start by preparing state $|00\rangle_L$ using the circuit shown in figure \ref{fig:circuits}(a). The results of measuring this state directly after preparation are shown in figure \ref{fig:prepzero}(a). The target states $0$ $(|00000\rangle)$ and $30$ $(|11110\rangle)$ are subject to readout errors, which have a larger effect on $30$. A re-normalized version of this data is shown here to illustrate that we succeed in preparing this state with $\simeq 92\%$ probability. The uncorrected data yields $91.1\%$ even-parity outcomes from the four data qubits. Breaking these results down by logical state gives $98.0\%$ population in the target state $|00\rangle_L$. The error falls almost entirely on $|01\rangle_L$, which corresponds to a $1.7\%$ error on the non-fault-tolerant gauge qubit $L_b$. The $0.1\%$ error exclusively on $L_a$ is an order of magnitude lower, and at a similar level as the $0.2\%$ logical two-qubit error resulting in $|11\rangle_L$. For the logical state preparation step, both of these small erroneous state populations are dominated by physical readout errors. 

\begin{figure}[htb]
\centering
\includegraphics[width=0.75\columnwidth]{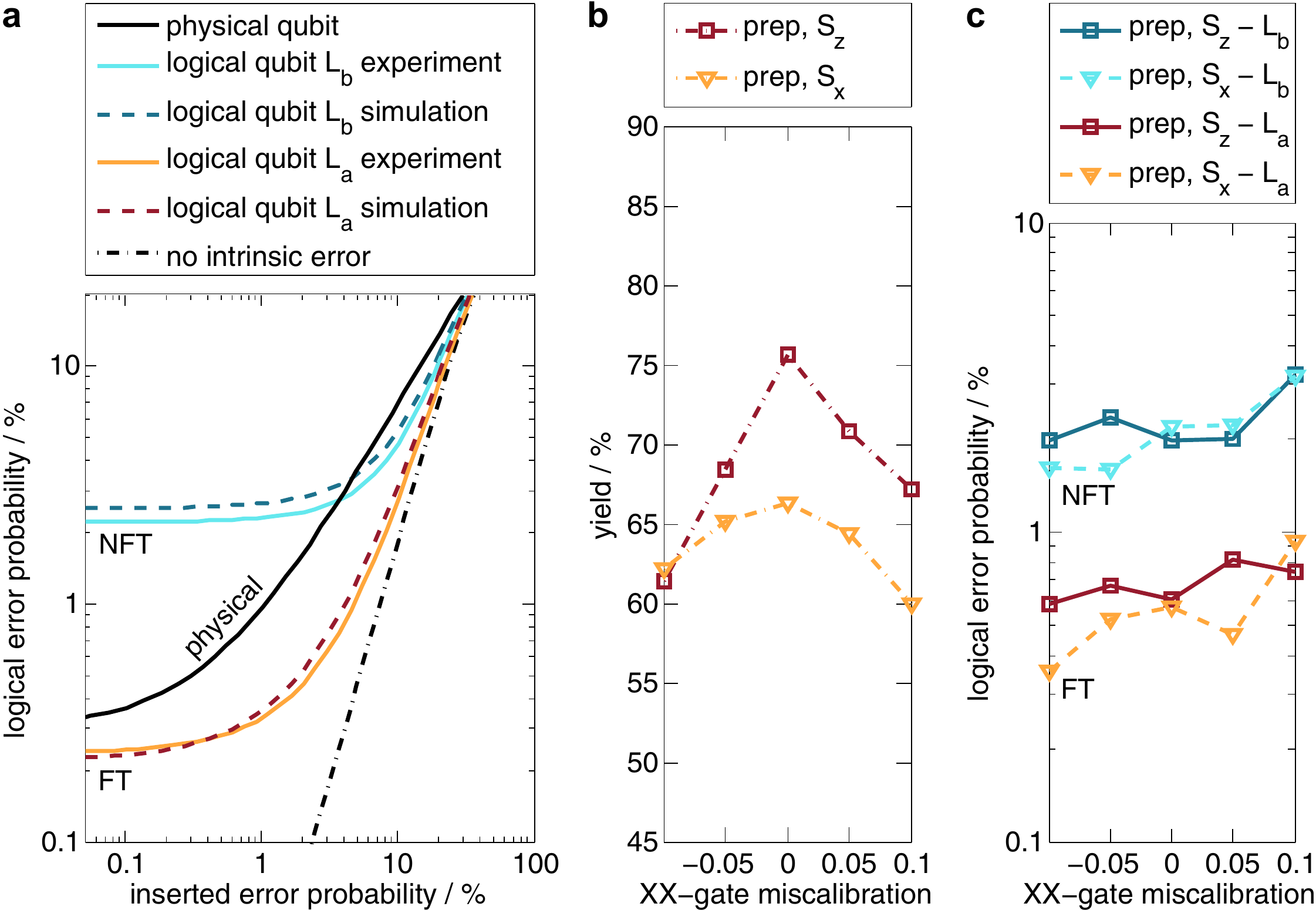} 
\caption{(a) Logical error probability under artificially introduced stochastic Pauli errors. We prepare state $|00\rangle_L$, introduce a specific error, and apply $S_z$ before readout. The parameter values for the curves (see Materials and Methods) corresponding to the two logical qubits are determined either experimentally (solid lines) or from simulation (dashed lines). The black curve shows the limiting theoretical case without intrinsic errors (see supplementary materials). At low added error rates, the intrinsic errors dominate and the fault-tolerant (FT) qubit $L_a$ starts about an order of magnitude below the non-fault-tolerant (NFT) qubit $L_b$. With increasing inserted error probability, the added Pauli errors become dominant and the $L_{a/b}$ curves converge and approach the theory curve without intrinsic error. The solid black line shows the error rate for a single physical qubit. $L_a$ results in a lower error across the entire range relative to the physical qubit while $L_b$ is lower for added errors $>4\%$. (b-c) Preparing $|00\rangle_L$ (prep) and measuring $S_{x/z}$ with purposefully miscalibrated two-qubit gates, known as XX-gates. A miscalibration of $\alpha$ means that the Bell state produced by the gate is imbalanced: $\sqrt{0.5-\alpha}|00\rangle+i\sqrt{0.5+\alpha}|11\rangle$. (c) shows the yields diminishing with miscalibration for the stabilizer measurements while the errors on the logical qubits presented in (b) stay similar, with $L_a$ errors about an order of magnitude lower than $L_b$ errors.}
\label{fig:miscal}
\end{figure}

With $|00\rangle_L$ thus prepared, we apply in turn the two stabilizers $S_z$ and $S_x$, shown in figure \ref{fig:circuits}(e, f) for non-demolition syndrome extraction. The results are shown in figure \ref{fig:prepzero}(b). Populations in the odd-numbered states reflect events where an error is detected by a stabilizer. The results of the logical states are similar, with $|00\rangle_L$ populations of $97.8\%$ and $97.1\%$, respectively, and the errors occurring predominantly in the non-fault-tolerant gauge qubit $L_b$. The errors on $L_a$ are $0.3\%$, similar to the error floor given by the $|11\rangle_L$ population, which is slightly higher than after mere state preparation due to the additional gates introduced by the stabilizers. $S_x$ introduces $X$-type errors in the system, which can be seen from a higher $L_b$ error. $S_z$ introduces $Z$-type errors, which do not affect $|00\rangle_L$. The opposite is true when applying the stabilizers to $|++\rangle$ instead. Table \ref{tb:summary} (see Supplementary materials) summarizes results for different logical states prepared with the circuits shown in figure \ref{fig:circuits}(a-d). The circuit elements that dominate the intrinsic errors in our system are the two-qubit gates. It is worth pointing out that after a circuit with $7$ CNOT gates, each of which introduces $~3-4\%$ infidelity, we get the correct answer $|00\rangle_L$ with $~97\%$ probability. The gauge qubit $L_b$ circuit failures occur at approximately the error rate of one two-qubit gate while $L_a$ errors are suppressed substantially below that level to $<1\%$. These results clearly show the power of fault-tolerant preparation and stabilizer measurement. The circuits succeed in discarding nearly all errors, but we pay a price as the yield is in the $65-75\%$ range. We must expect to discard around half of the runs when measuring both stabilizers. The yield is higher for preparation of logical states without syndrome measurements because there are fewer gates to introduce error and only a single selection step.

The [[4,2,2]] code allows transversal operations, i.e. single-qubit logical gates that are generated by applying single-qubit physical gates. To show an example of this, we prepare $|00\rangle_L$ followed by the logical $X_a X_b$-operation consisting of $X$-gates on physical qubits $2$ and $3$. This gives $|11\rangle_L$, on which we apply the $S_x$ stabilizer followed by readout. The yield is $73.3\%$ and the logical state populations are $|00\rangle_L$: $0.4\%$, $|01\rangle_L$: $0.3\%$, $|10\rangle_L$: $2.8\%$, $|11\rangle_L$: $96.5\%$. Apart from surpassing $L_b$ as before, the $L_a$ error of $0.7\%$ also outperforms the physical qubit. We find that after an $X$-gate the correct state of a physical qubit is measured with $98.8\%$ fidelity, nearly a factor of two worse than $L_a$. The infidelity in this case is dominated by the single-qubit detection error of $0.9\%$ for $|1\rangle$, which the code successfully suppresses in $L_a$.

In order to further investigate the robustness of the code and the fault tolerance of its logical qubits, we add two kinds of error to the system. Firstly, we deliberately introduce single- and two-qubit Pauli errors and study how errors on $L_{a}$ and $L_b$ scale with increasing physical qubit errors. Instead of trying to reproduce a stochastic error channel, which can be tedious for low error rates \cite{Bravyi13}, we sample the various error configurations and then multiply them by their respective statistical importance to get a logical error probability (see supplementary materials). We further compare our experimental results to an exact simulation with optimized error parameters (see supplementary materials). The results are shown in figure \ref{fig:miscal}(a). The clear separation between the two logical qubits is persistent until they converge above $20$\% introduced error and approach the curve for the theoretical case without intrinsic errors. 
In this example, a physical qubit prepared in state $|0\rangle$ is outperformed by logical qubit $L_a$ over the entire range and by $L_b$ above $4\%$ added error (solid black line in figure \ref{fig:miscal}(a)). For state preparations $|-\rangle$ and $|1\rangle$, $L_a$ also outperforms the physical error based on circuits of preparation and measurement, while for $|+\rangle$ the errors are consistent within statistical uncertainty (see supplementary materials table \ref{tb:summary}). 

Secondly, we run the $|00\rangle_L$ data with purposefully miscalibrated $2$-qubit gates. The results are shown in figure \ref{fig:miscal}(c). The error gap of nearly an order of magnitude between $L_a$ and $L_b$ persists over a wide range of calibration errors, which are absorbed into a reduced yield as shown in figure \ref{fig:miscal}(b). This proves that the code succeeds in protecting qubit $L_a$ against intrinsic systematic errors. 

We note that the [[4,2,2]] code is relevant beyond its limited immediate application as an error detection code. It forms the base encoding layer of the high-threshold Knill C4/C6 code \cite{Knill05} and of a recent proposal for a topological code \cite{Terhal16}, and it is equivalent to one face of the distance-$3$ color code \cite{Bombin06} or the Steane code \cite{Steane96}. It is robust to the high levels of intrinsic errors present in current realizations of quantum computers, paving the way towards error-corrected quantum computations on a larger scale.


\section*{Acknowledgments}
We thank J. Kim and D. Maslov for helpful discussions, and D. Gottesman for useful comments. This work was supported by the ARO with funds from the IARPA LogiQ program, the AFOSR MURI program on Quantum Measurement and Verification, and the NSF Physics Frontier Center at JQI. Author contributions. N.M.L., M.G., K.A.L., C.F., S.D., K.R.B., and C.M. all contributed to the experimental design, construction, data collection and analysis of this experiment. All authors contributed to this manuscript. Declaration of competing financial interests: C.M. is a founding scientist of IonQ, Inc. List of Supplementary materials: Materials and Methods, Table 1.

\newpage
\clearpage

\subsection*{Supplementary materials}
\subsubsection*{Materials and Methods}

\noindent
{\bf Experimental system} The experiment is performed on a quantum computer consisting of a chain of five single $^{171}$Yb$^+$ ions confined in a Paul trap and laser cooled near the motional ground state. Each ion provides one physical qubit in the form of a pair of states in the hyperfine-split $^2S_{1/2}$ ground level with an energy difference of $12.642821\:$GHz, which is magnetic field independent to first order. This so-called ``atomic clock'' qubit has a typical coherence time of $~0.5\:$s, which can be straightforwardly extended by suppressing magnetic field noise. All qubits are collectively initialized by optical pumping and measured via state-dependent fluorescence detection \cite{Olmschenk07}. Each ion is mapped to a distinct channel of a photomultiplier tube (PMT) array. Its state can be detected with $99.4(1)\%$ average fidelity, while a $5$-qubit state is read out with $95.7(1)\%$ average fidelity, limited by channel-to-channel crosstalk. Qubit manipulation is achieved by applying two Raman beams from a single $355\:$nm mode-locked laser, which form beat notes near the qubit frequency. The first Raman beam is a global beam applied to the entire chain, while the second is split into individual addressing beams, each of which can be switched independently to target any single qubit \cite{Debnath16}. Single qubit gates are generated by driving resonant Rabi rotations of defined phase, amplitude, and duration. Two-qubit gates (so-called XX-gates) are realized by illuminating two ions with beat-note frequencies near the motional sidebands and creating an effective spin-spin (Ising) interaction via transient entanglement between the state of two ions and all modes of motion \cite{Molmer99, Solano99, Milburn00}. To ensure that the motion is left disentangled from the qubit states at the end of the interaction, we employ a pulse shaping scheme by modulating the amplitude of the global beam \cite{Zhu06eur,Choi14}. 

\noindent
{\bf Artificial stochastic errors} To analyze how the code copes with artificially introduced stochastic errors, we prepare logical state $|00\rangle_L$ and add a specific Pauli error, e.g. $I\otimes X \otimes Y \otimes I$. We then apply the $S_z$ stabilizer and measure the state. We repeat this for different error configurations $\varepsilon$. The error probability $p$ on a physical qubit corresponds to an $X$, $Y$, or $Z$ error, each occurring with probability $p/3$. How many errors appear in a particular error configuration is given by its weight $w$, and its probability of occurrence or statistical importance is $p_o=(p/3)^w(1-p)^{4-w}$. The probability of a logical error is given by
\begin{equation}
\label{eq:logicalerror}
p_L=\frac{\sum_{\varepsilon}{p_o(\varepsilon) \cdot p_a(\varepsilon) \cdot p_f(\varepsilon)}}{\sum_{\varepsilon}{p_o(\varepsilon) \cdot p_a(\varepsilon)}}
\end{equation}
The sum runs over all error configurations. $p_a$ is the yield, i.e. the probability that a run is accepted, and $p_f$ is the probability of failure after post-selection, i.e. the probability that an accepted run suffers a logical error. The dividend is the number of accepted runs with a logical error, while the divisor is the number of accepted runs (both divided by the total number of runs). The parameters $p_a$ and $p_f$ are found either from experiment or simulation (see figure \ref{fig:miscal}(a)). Out of the total number of error configurations $n(w) = w^3 {4\choose w}$. We cover the error configurations of weight $0$ and $1$ exhaustively. For the $w=2$ subset, we only sample $27$ representative configurations out of the total $54$ and double their weight. The weight-$3$ and -$4$ subsets are not sampled and their logical error rates set to zero, since their statistical importance is significant only at very high added error rates. 
In the limit of no intrinsic errors, i.e. perfect gates, preparation, and measurement, both logical qubits have the same error rate under this model (dash-dotted line in figure \ref{fig:miscal}(a)). We find this error rate from eq. \ref{eq:logicalerror} by counting accepted error configurations with $w\leq2$ (denominator) and checking which of those cause an error (numerator).
\begin{equation}
\label{eq:perfect}
p_L^*=\frac{16(1-p)^2(p/3)^2}{(1-p)^4 + 4 (1-p)^3 p/3 + 30(1-p)^2(p/3)^2}
\end{equation}

The dashed curves in figure \ref{fig:miscal}(a) are obtained by performing a full density matrix simulation of the $5$-qubit circuits. We use a simplified error model to emulate experimental errors. The model has $3$ independent parameters corresponding to errors associated with over- or under-rotations after (1) single-qubit and (2) two-qubit gates, and (3) phase errors caused by Stark shifts. An experimentally found state-transfer matrix is used to take state-preparation and detection errors, including crosstalk, into account. We then optimize the model over the parameter space to minimize the difference between the final state populations of the experimental and simulated circuits. The resulting values for the error rates are $0.50\%$, $1.0\%$, and $1.4\%$, respectively.

The physical error curve in figure \ref{fig:miscal}(a) is the straight line $p_p = r + (2/3\;F_x)p$ where $r=0.003$ is the readout error for a physical qubit in state $|0\rangle$. The slope is $2/3 \; F_x$ since one in three added errors is a $Z$-type error which does not affect $|0\rangle$, and $F_x=0.997$ is the success probability of a physical spin flip operation.

\newpage
\subsection*{Logical state preparation data}
\begin{table}[ht!]
\centering
\begin{tabular}{|l|r|r|r|r|r|r|}
\hline
 & & \multicolumn{4}{c|}{ }  & \multicolumn{1}{l|}{meas.}\\
 & & \multicolumn{4}{c|}{meas. logical state $|L_a L_b\rangle$} & \multicolumn{1}{l|}{basis} \\
  &  &  $|00\rangle$ & $|01\rangle$ & $|10\rangle$ & $|11\rangle$ & Z\\
	  & yield  &  $|++\rangle$ & $|+-\rangle$ & $|-+\rangle$ & $|--\rangle$ & X\\
\hline
\hline
$|00\rangle_L$ & 91.1  & 98.0 & 1.7 & 0.1 & 0.2 & Z \\
\hline
$|00\rangle_L S_z$ & 77.8 & 97.8 & 1.7 & 0.2 & 0.3 & Z \\
\hline
$|00\rangle_L S_x$ & 65.2 & 97.1 & 2.4 & 0.2 & 0.3 & Z \\
\hline
$|++\rangle_L$ & 91.1 & 94.8 & 3.9 & 0.2 & 0.2 & X \\ 
\hline
$|++\rangle_L S_z$ & 68.2 & 93.0 & 4.2 & 1.3 & 1.5 & X \\ 
\hline
$|++\rangle_L S_x$ & 72.1 & 94.3 & 4.5 & 0.5 & 0.7 & X \\ 
\hline
$|-1\rangle_L$ & 90.1 & 0.2 & 50.5 & 0.1 & 49.2 & Z \\ 
\hline
$|-1\rangle_L$ & 87.0 & 0.3 & 0.3 & 50.4 & 48.9 & X \\ 
\hline
$|-1\rangle_L S_z$ & 79.9 & 0.2 & 50.0 & 0.1 & 49.7 & Z \\ 
\hline
$|-1\rangle_L S_z$ & 75.5 & 0.4 & 0.3 & 50.0 & 49.2 & X \\ 
\hline
$|-1\rangle_L S_x$ & 72.1 & 0.6 & 50.2 & 0.5 & 48.7 & Z \\ 
\hline
$|-1\rangle_L S_x$ & 76.2 & 0.4 & 0.4 & 50.0 & 49.2 & X \\ 
\hline
$|0+\rangle_L$ & 93.2 & 47.4 & 52.5 & 0.06 & 0.05 & Z \\ 
\hline
$|0+\rangle_L$ & 92.4 & 50.0 & 0.04 & 49.8 & 0.09 & X \\ 
\hline
$|0+\rangle_L S_z$ & 81.6 & 48.3 & 51.3 & 0.2 & 0.2 & Z \\ 
\hline
$|0+\rangle_L S_z$ & 68.5 & 47.1 & 2.4 & 47.4 & 3.1 & X \\ 
\hline
$|0+\rangle_L S_x$ & 72.0 & 48.3 & 51.5 & 0.2 & 0.1 & Z \\ 
\hline
$|0+\rangle_L S_x$ & 70.9 & 49.4 & 0.4 & 49.7 & 0.5 & X \\ 
\hline
$|11\rangle_L S_z$ & 73.3 & 0.4 & 0.3 & 2.8 & 96.5 & Z \\ 
\hline
\end{tabular}
\caption{Probability distributions (in percent) of measured logical states $|L_a L_b\rangle$ for various prepared logical states in each row, with and without stabilizers $S_x$ or $S_z$ applied. The measurement basis is shown in the last column. The logical states are $|00\rangle_L \ldots |11\rangle_L$, measured in the $Z$-basis, and $|++\rangle_L\ldots |--\rangle_L$, measured in the $X$-basis. The very low error probability on the first logical qubit $L_a$ compared to $L_b$ shows clearly the action of its fault-tolerant construction. We run every circuit $5000-6000$ times resulting in a statistical uncertainty of $0.1\%$ on the numbers given. The results without stabilizer show the number of rejected runs from the parity check on the data qubits (typically $\sim 8\%$) while the additional discard (typically $\sim 20\%$) is due to the ancilla result. The physical errors for state preparation and measurement are $0.3\%$ for states $|0\rangle$ and $|+\rangle$, and $1.2\%$ for states $|1\rangle$ and $|-\rangle$.}
\label{tb:summary}
\end{table}


\clearpage
\begin{thebibliography}{10}

\bibitem{Calderbank96}
A.~R. Calderbank, P.~W. Shor, {\it Phys. Rev. A\/} {\bf 54}, 1098 (1996).

\bibitem{Steane96}
A.~M. Steane, {\it Proc. Roy. Soc. A\/} {\bf 452}, 2551 (1996).

\bibitem{Preskill98}
J.~Preskill, {\it Proc. Roy. Soc. A\/} {\bf 454}, 385 (1998).

\bibitem{Nielsen11}
M.~A. Nielsen, I.~L. Chuang, {\it Quantum Computation and Quantum Information:
  10th Anniversary Edition\/} (Cambridge University Press, New York, NY, USA,
  2011), 10th edn.

\bibitem{Gottesman09}
D.~Gottesman, {\it Proceedings of Symposia in Applied Mathematics\/}, S.~J.~L.
  Jr., ed. (American Mathematical Society, 2009), vol.~68, pp. 13--59.

\bibitem{Chiaverini04}
J.~Chiaverini, {\it et~al.\/}, {\it Nature\/} {\bf 432}, 602 (2004).

\bibitem{Schindler11}
P.~Schindler, {\it et~al.\/}, {\it Science\/} {\bf 332}, 1059 (2011).

\bibitem{Lanyon13}
B.~P. Lanyon, {\it et~al.\/}, {\it Phys. Rev. Lett.\/} {\bf 111}, 210501
  (2013).

\bibitem{Waldherr14}
G.~Waldherr, {\it et~al.\/}, {\it Nature\/} {\bf 506}, 204 (2014).

\bibitem{Nigg14}
D.~Nigg, {\it et~al.\/}, {\it Science\/} {\bf 345}, 302 (2014).

\bibitem{Kelly15}
J.~Kelly, {\it et~al.\/}, {\it Nature\/} {\bf 519}, 66 (2015).

\bibitem{Corcoles2015}
A.~D. C\'{o}rcoles, {\it et~al.\/}, {\it Nature Comm.\/} {\bf 6}, 7979 (2015).

\bibitem{Cramer16}
J.~Cramer, {\it et~al.\/}, {\it Nature Comm.\/} {\bf 7}, 11526 (2016).

\bibitem{Ofek16}
N.~Ofek, {\it et~al.\/}, {\it Nature\/} {\bf 536}, 441 (2016).

\bibitem{Grassl97}
M.~Grassl, T.~Beth, T.~Pellizzari, {\it Phys. Rev. A\/} {\bf 56}, 33 (1997).

\bibitem{Gottesman16}
D.~Gottesman, {\it arXiv:1610.03507\/}  (2016).

\bibitem{Lu08}
C.-Y. Lu, {\it et~al.\/}, {\it Proc. Natl. Acad. Sci. USA\/} {\bf 105}, 11050
  (2008).

\bibitem{Bell14}
B.~A. Bell, {\it et~al.\/}, {\it Nature Comm.\/} {\bf 5}, 3658 (2014).

\bibitem{Cross09}
A.~W. Cross, D.~P. Divincenzo, B.~M. Terhal, {\it Quantum Inf. Comput.\/} {\bf
  9}, 541 (2009).

\bibitem{Bacon06}
D.~Bacon, {\it Phys. Rev. A\/} {\bf 73}, 012340 (2006).

\bibitem{Shor95}
P.~W. Shor, {\it Phys. Rev. A\/} {\bf 52}, R2493 (1995).

\bibitem{Napp13}
J.~Napp, J.~Preskill, {\it Quantum Inf. Comput.\/} {\bf 13}, 490 (2013).

\bibitem{Takita16}
M.~Takita, {\it et~al.\/}, {\it Phys. Rev. Lett.\/} {\bf 117}, 210505 (2016).

\bibitem{Debnath16}
S.~Debnath, {\it et~al.\/}, {\it Nature\/} {\bf 536}, 63 (2016).

\bibitem{Bravyi13}
S.~Bravyi, A.~Vargo, {\it Phys. Rev. A\/} {\bf 88}, 062308 (2013).

\bibitem{Knill05}
E.~Knill, {\it Nature\/} {\bf 434}, 39 (2005).

\bibitem{Terhal16}
B.~Criger, B.~Terhal, {\it arXiv:1604.04062\/}  (2016).

\bibitem{Bombin06}
H.~Bombin, M.~A. Martin-Delgado, {\it Phys. Rev. Lett.\/} {\bf 97}, 180501
  (2006).

\bibitem{Olmschenk07}
S.~Olmschenk, {\it et~al.\/}, {\it Phys. Rev. A\/} {\bf 76}, 052314 (2007).

\bibitem{Molmer99}
K.~M\o{}lmer, A.~S\o{}rensen, {\it Phys. Rev. Lett.\/} {\bf 82}, 1835 (1999).

\bibitem{Solano99}
E.~Solano, R.~L. de~Matos~Filho, N.~Zagury, {\it Phys. Rev. A\/} {\bf 59},
  R2539 (1999).

\bibitem{Milburn00}
G.~Milburn, S.~Schneider, D.~James, {\it Fortschritte der Physik\/} {\bf 48},
  801 (2000).

\bibitem{Zhu06eur}
S.-L. Zhu, C.~Monroe, L.-M. Duan, {\it Europhys Lett.\/} {\bf 73}, 485 (2006).

\bibitem{Choi14}
T.~Choi, {\it et~al.\/}, {\it Phys. Rev. Lett.\/} {\bf 112}, 190502 (2014).

\end{thebibliography}
\end{document}